%% file: frinchaboy.tex
\documentclass[graybox]{svmult}

\usepackage{mathptmx}       
\usepackage{helvet}         
\usepackage{courier}        
\usepackage{type1cm}        
\usepackage{makeidx}         
\usepackage{graphicx}        
\usepackage{multicol}        
\usepackage[bottom]{footmisc}

\makeindex             

\begin{document}

\title*{SDSS-III/APOGEE: Detailed Abundances of Galactic Star Clusters}
\author{Peter M Frinchaboy, Gail Zasowski, Kelly Jackson, Jennifer
  A. Johnson, Steven R. Majewski, Matthew Shetrone, Aaron Rocha, for the SDSS-III collaboration}
\authorrunning{P.M. Frinchaboy, for the SDSS-III collaboration} 
\institute{Peter M. Frinchaboy, Kelly Jackson, Aaron Rocha \at Department of Physics \& Astronomy, Texas Christian University,
TCU Box 298840, Fort Worth, TX 76129, USA \email{p.frinchaboy@tcu.edu,
  kelly.m.jackson@tcu.edu, a.a.rocha@tcu.edu}
\and Gail Zasowski, Steven R. Majewski \at Department of Astronomy, University of Virginia, 
P.O. Box 400325, Charlottesville, VA 22904-4325, USA
\email{gz2n@virginia.edu, srm4n@virginia.edu}
\and Jennifer A. Johnson \at Department of Astronomy, The Ohio State
University, 140 West 18th Avenue, Columbus, OH 43210, USA \email{jaj@astronomy.ohio-state.edu}
\and  Matthew Shetrone\at McDonald Observatory, University of Texas at
Austin, HC75 Box 1337-MCD, Fort Davis, TX 79734, USA \email{shetrone@astro.as.utexas.edu}}
%
%
\maketitle
\vskip-1in
\abstract*{ The Sloan Digital Sky Survey III/Apache Point Observatory
  Galactic Evolution Experiment (SDSS-III/APOGEE) is a large-scale
  spectroscopic survey of Galactic stars and star clusters. The
  SDSS-III/APOGEE survey is designed to produce high-$S/N$, $R =$
  27,500--31,000 spectra that cover a wavelength range of 1.51 to 1.68
  microns. By utilizing APOGEE's excellent kinematics (error $\le$ 0.5
  km s$^{-1}$) and abundances (errors $\le$ 0.1 dex), we will be able to
study star cluster kinematics and chemical properties in detail.  Over
the course of the 3-year survey beginning in 2011, APOGEE will target
25--30 key open and globular clusters.  In addition, the large area
coverage of the SDSS focal plane will also allow us to target stars in
100--200 additional star clusters during the main survey observations.
We present the strength of APOGEE for both open and globular star
cluster studies and the methods of identifying probable clusters members utilizing 2MASS and IRAC/WISE data.}
\abstract{ The Sloan Digital Sky Survey III/Apache Point Observatory
  Galactic Evolution Experiment (SDSS-III/APOGEE) is a large-scale
  spectroscopic survey of Galactic stars and star clusters. The
  SDSS-III/APOGEE survey is designed to produce high-$S/N$, $R =$
  27,500--31,000 spectra that cover a wavelength range of 1.51 to 1.68
  microns. By utilizing APOGEE's excellent kinematics (error $\le$ 0.5
  km s$^{-1}$) and abundances (errors $\le$ 0.1 dex), we will be able to
study star cluster kinematics and chemical properties in detail.  Over
the course of the 3-year survey beginning in 2011, APOGEE will target
25--30 key open and globular clusters.  In addition, the large area
coverage of the SDSS focal plane will also allow us to target stars in
100--200 additional star clusters during the main survey observations.
We present the strength of APOGEE for both open and globular star
cluster studies and the methods of identifying probable clusters members utilizing 2MASS and IRAC/WISE data.}

\section{Introduction}
Star clusters represent a key tracer for the dynamical and chemical
evolution of galaxies.  The one galaxy for which we can investigate
{\it in detail} is our own Milky Way galaxy.  While there have
been wide ranging studies of Galactic star clusters, there remains a 
key problem with studying Galactic evolution: lack of large uniform
samples.  For photometric studies, this is starting to be
possible with large-area surveys (e.g., 2MASS, SDSS-I, VVV, Skymapper, 
LSST).  Large kinematic samples have begun to be derived utilizing proper motions (e.g.,
\cite{glush,baum,dias01,dias02,dias06}) and radial velocities
(e.g., \cite{pm08}).  However,
high-resolution spectroscopy studies, yielding high accuracy radial velocities and 
detailed chemical abundances, are still limited to small sets of
clusters\footnote{High-resolution studies of up to ten open clusters are starting to
be published using CTIO/WIYN Hydra (e.g., \cite{jacobson09,friel10}) 
and VLT/Flames+UVES (e.g., \cite{sestito06,bragaglia08,sestito08})}.  
The soon to be commissioned Sloan Digital Sky Survey III/Apache
Point Observatory Galactic Evolution Experiment (SDSS-III/APOGEE) will 
provide for uniform data and analysis for a survey of up to 200 star clusters.

\section{SDSS-III/APOGEE}
The SDSS-III/APOGEE project is three-year high-resolution
spectroscopic survey that will cover all major Galactic populations (thin disk,
thick disk, bulge/bar, and halo).  The project utilizes a new
300-fiber-fed  $H$-band (1.51 to 1.68 $\mu$m) spectrograph \cite{apogeeinst}.
The spectrograph will yield $R$ = 27,500-31,000 spectra with $S/N \sim
100$ per pixel for stars with $H = 12.3$.  The goal of the survey is
to derive precision radial velocities ($\sigma_v \le 0.5$ km s$^{-1}$)
and abundances ($\sigma_{[X/Fe]} \le 0.1$ dex) for $\sim 100,000$
stars.  The survey has
planned to study 15 different elements (including Fe, C, N, O,
$\alpha$-elements, odd-$Z$ elements, iron peak elements, possibly even
neutron capture).  The survey will coordinate observations with
another SDSS-III survey, the Multi-object APO Radial Velocity
Exoplanet Large-area Survey (SDSS-III/MARVELS) survey, which will necessitate
that 75\% of the observing time be spent in 58 key fields (which
contain $\sim$60 star clusters).
The total APOGEE survey plans to target 220--230 unique field
centers covering $\sim 1200$ deg$^2$ of the sky.

\subsection{APOGEE Calibration Clusters}

One area of APOGEE science and calibration is the study of key
star clusters listed in Table~\ref{tab:1}, with parameters taken from
Harris catalog \cite{harris} for globular clusters and the Dias
catalog \cite{diascat} for the open clusters.  These key clusters
will have at least one 7 deg$^2$ plate configuration, up to 250 fibers,  
dedicated to likely cluster member stars.  For some of the open
clusters (e.g., M67, NGC 188, NGC 6819) we have kinematic membership
and binary information available from the WIYN Open Cluster Study
(WOCS; \cite{mathieu,geller,hole})

We will explicitly target many stars which have already been observed
at high resolution ($R > 30,000$) in the optical or near infrared to
be used to compare SDSS-III/APOGEE to other high-resolution studies.
For these ``calibration'' cluster targets, we will obtain large
numbers of members that will be used to
fully characterize the clusters bulk chemical properties, but also
allow science ranging for looking for abundance variations on the
individual element level to investigating isotopic abundance
variations as keys to understanding evolution along the red giant branch.

\begin{table}
\caption{APOGEE candidate calibration clusters }
\label{tab:1}       
\begin{tabular}{p{2cm}p{1cm} c r r r r r r r r r r r r r r r }
\hline\noalign{\smallskip}
  Name &  & Type &&  Diam && [Fe/H] &&  $\sigma_{[Fe/H]}$  && Age (Yr) && Log(Age) && Dist(pc) \\
\noalign{\smallskip}\svhline\noalign{\smallskip}
   NGC 188      &      &    Open Cl    &&    17' &&    -0.01  &&      0.09     &&      4.2 Gyr &&      9.632   &&        2047  \\
   Pleiades     &      &    Open Cl    &&   110' &&    -0.03  &&      0.06     &&      135 Myr &&      8.131   &&         150  \\
   Hyades       &  M45 &    Open Cl    &&   330' &&    +0.13  &&      0.06     &&      787 Myr &&      8.896   &&          45  \\
   NGC 2168     &  M35 &    Open Cl    &&    25' &&    -0.16  &&      0.09     &&       95 Myr &&      7.979   &&         816  \\
   NGC 2243     &      &    Open Cl    &&     5' &&    -0.49  &&      0.05     &&      1.1 Gyr &&      9.032   &&        4458  \\
   Melotte 71   &      &    Open Cl    &&     7' &&    -0.30  &&      0.06     &&      235 Myr &&      8.371   &&        3154  \\
   NGC 2420     &      &    Open Cl    &&     5' &&    -0.40  &&               &&      2.8 Gyr &&      9.45    &&        2290  \\
   NGC 2682     &  M67 &    Open Cl    &&    25' &&    -0.15  &&      0.05     &&      2.5 Gyr &&      9.409   &&         908  \\
   NGC 6171     & M107 &    Globular   &&    17' &&    -0.90  &&      0.10     &&         GC   &&      GC      &&        6400  \\
   NGC 6205     & M13  &    Globular   &&    25' &&    -1.51  &&      0.10     &&         GC   &&      GC      &&        7700  \\
   IC 4725      & M25  &    Open Cl    &&    29' &&    +0.17  &&      0.06     &&       92 Myr &&      7.965   &&         620  \\
   NGC 6791     &      &    Open Cl    &&    10' &&    +0.35  &&      0.02     &&      4.4 Gyr &&      9.643   &&        5853  \\
   NGC 6819     &      &    Open Cl    &&     5' &&    +0.07  &&               &&      3.1 Gyr &&      9.490   &&        2360  \\
   NGC 6838     & M71  &    Globular   &&     9' &&    -0.79  &&      0.10     &&         GC   &&      GC      &&        6700  \\
   NGC 7078     & M15  &    Globular   &&    21' &&    -2.20  &&      0.10     &&         GC   &&      GC      &&       10300  \\
   NGC 7089     &  M2  &    Globular   &&    21' &&    -1.62  &&               &&         GC   &&      GC      &&       11500  \\
   NGC 7789     &      &    Open Cl    &&    25' &&    -0.20  &&               &&      1.7 Gyr &&      9.230   &&        1820  \\
\noalign{\smallskip}\hline\noalign{\smallskip}
\end{tabular}
\end{table}

\section{APOGEE Candidate Cluster Analysis}
\subsection{Photometry Analysis}

The only all-sky photometry that will be available for APOGEE
targeting will come from two sources Two-Micron All-Sky Survey (2MASS $J H K_S$; \cite{2mass}) and the soon to be
released {\it Wide-field Infrared Survey Explorer} ({\it WISE}; \cite{wise}) mission.
However, since many of the targets will be in the Galactic midplane,
and WISE has poor resolution ($\sim 6$ arcsec), we will also
supplement our data set with {\it Spitzer}/IRAC Galactic Legacy
Infrared Mid-Plane Survey Extraordinaire (GLIMPSE I, II, 3D, 360;
\cite{glimpse})
surveys, which provide better resolution in the needed [3.6] and [4.5]
micron bandpasses.

\begin{figure}[t]
\includegraphics[angle=90,scale=.59]{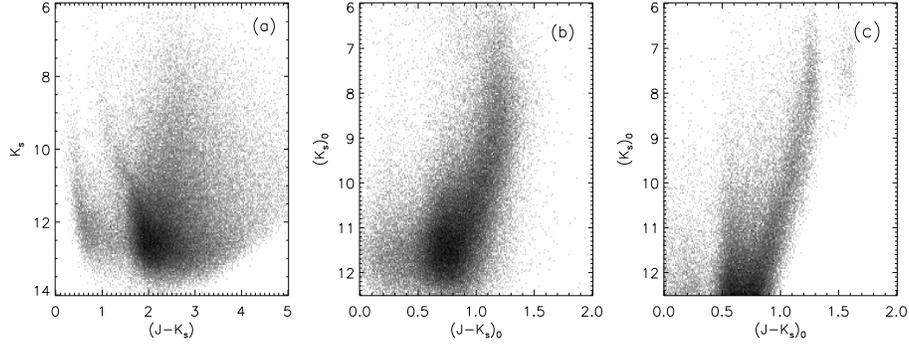}
%
%
\caption{Demostration of the RJCE technique.  a) Raw 2MASS CMD for a 4 deg$^2$
  field at ($l,b$) = (42,0).  b) Field dereddened using the RJCE
  technique.  c) TRILEGAL \cite{trilegal} simulation of the Galaxy for a field at ($l,b$) = (42,0).}
\label{fig:rjce}       
\end{figure}

These surveys will provide five-band photometry $JHK_S$[3.6][4.5] data
which allow us to derive star-by-star extinctions utilizing the
Rayleigh-Jeans Color Excess (RJCE)
method (see Figure~\ref{fig:rjce}), fully described in Majewski et al. \cite{RJCE}.  The RJCE
method allows us to explicitly determine the $A_{K_S}$ extinction to
each star by using the observed $H-$[4.5] color that is nearly
constant for a large range of common spectral types.
The ability to derive extinctions and correct to relative distance
ranges is essential for a survey of the Galactic plane and
bulge, and we can utilize this tool to isolate potential cluster stars
from the field population. 

\subsection{Cluster Analysis}
\begin{figure}[ht!]
\includegraphics[scale=.59]{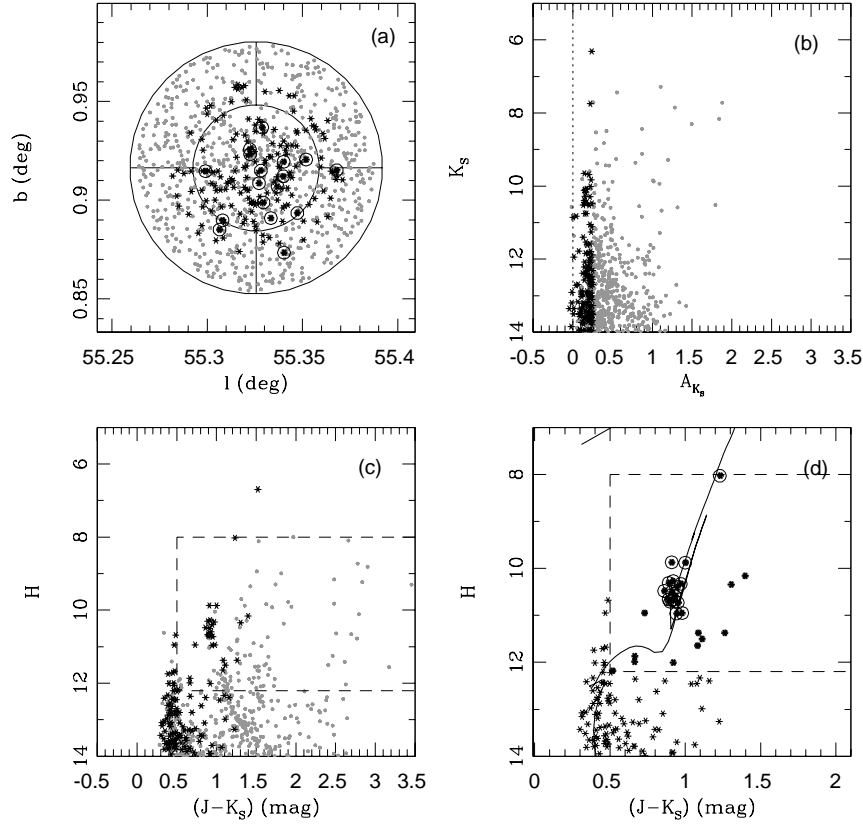}
%
%
\caption{Sample analysis for the cluster NGC 6802 utilizing
  2MASS+GLIMPSE data. a) Galactic latitude and longitude  for all stars (gray) within the $2R_{cl}$ area to
  be analyzed, stars selected to be likely members from the photometry
  extinction analysis are shown in black.  Prime APOGEE targets are
  circled. b) Distribution of $A_{K_s}$ for all stars in the NGC 6802 
  sample area, black points denote stars with $1.1 R_{cl}$ within the
  determined mean cluster $A_{K_s}$ range.  c) Color-magnitude diagram
  (CMD) for all stars in the analysis area (gray). The dashed box
  denotes the SDSS-III/APOGEE target selection region.  Black points
  denote stars selected as likely members from their $A_{K_s}$. d) CMD of only likely
  cluster members overplotted with the Padova Isochrone
  \cite{isochrone} using the
  clusters parameters from \cite{diascat}.  Circled stars
  denote identified high-probability stars for APOGEE target selection
  (also see the on-sky distribution in (a))
}
\label{fig:cl}       
\end{figure}

In order to distinguish and isolate star clusters from foreground and
background contamination, we utilize the $A_{K_S}$ values derived from
the RJCE technique described above.  We isolate a region of
approximately twice the clusters catalog radius \cite{diascat} and
divide it into 5 regions (see Figure~\ref{fig:cl}a).  We utilize four
``background'' regions and the cluster region (radius = $R_{Dias}$).
The background is divided in order to account for dust clouds,
clusters near the edge of the GLIMPSE survey reigion, and any other source of
background variability.
We difference the mean field/background star numbers to the ``cluster'' star numbers within a given $A_{K_s}$
range, and scan this range across all available $A_{K_s}$ values that
have at least 15 stars (Mean field + cluster stars $\ge 15$; see Figure~\ref{fig:cl}b). The window of
extinction with the highest concentration of stars within the inner
radius will reveal the cluster (Figure~\ref{fig:cl}c \& d). 
We then work to optimize the cluster isolation surveying a grid of $A_{K_s}$ width, $A_{K_s}$
stepsize, and allowed $\sigma_{A_{K_s}}$ values. 

\subsection{Preliminary Results}

%

We present a first analysis for the cluster NGC 6802 to demonstrate
the technique, shown in Figure~\ref{fig:cl}.
Figure~\ref{fig:cl}a first shows the area explored by our analysis in Galactic latitude and longitude.
As desribed above, we selected likely cluster members utilizing the
$A_{K_s}$ as shown in Figure~\ref{fig:cl}b.  For NGC 6802 we find a
low, but non-negligible extinction or reddening to the cluster.
A color magnitude diagram (CMD) of the clusters (Figure~\ref{fig:cl}c)
is generated which highlights the member stars with $A_{K_s}$ values
within the window of extinction, where the dashed box
in the center denotes the area where the upcoming SDSS-III/APOGEE
project will be targeting ($8.0 < H < 12.3$ and $J-K_S \ge 0.5$). 
Finally, we compare our ``cleaned'' cluster CMD to the Padova 
isochrone utilizing catalog values \cite{diascat} for NGC 6802
and find a good match.  By comparing the CMD with isochrone values,
when available, we are able to isolate candidate cluster stars with a
high probablility for membership.  The APOGEE project requires this
cleaning for most clusters for two reasons: 1) most open clusters are
found a low Galactic latitude and thereby are heavily contaminated
with field stars. 2) Due to the large SDSS telescope field of view \cite{sdsstel},
the minimum fiber-to-fiber distance is fairly large $\ge 1$ arcmin,
which only allows for the targeting of a handfull of stars ($\sim
5-10$) per cluster for the most poorly studied, distant,
and reddened clusters.

\begin{acknowledgement}
SDSS-III is managed by the Astrophysical Research Consortium for the
Participating Institutions of the SDSS-III Collaboration including the
University of Arizona, the Brazilian Participation Group, Brookhaven
National Laboratory, University of Cambridge, University of Florida,
the French Participation Group, the German Participation Group, the
Instituto de Astrofisica de Canarias, the Michigan State/Notre
Dame/JINA Participation Group, Johns Hopkins University, Lawrence
Berkeley National Laboratory, Max Planck Institute for Astrophysics,
New Mexico State University, New York University, the Ohio State
University, University of Portsmouth, Princeton University, University
of Tokyo, the University of Utah, Vanderbilt University, University of
Virginia, University of Washington, and Yale University.   This work
is based in part on observations made with the Spitzer Space
Telescope, which is operated by the Jet Propulsion Laboratory,
California Institute of Technology under a contract with NASA.
Support for this work was provided by NASA through an award 
issued by JPL/Caltech.  Additionally KJ was supported by an NSF REU
grant (NSF 0851558) and funding from Texas
Christian University, including a Science and Engineering Research Center (TCU-SERC) grant.

\end{acknowledgement}
%


\input{frinchaboy_ref}

\end{document}

%% file: frinchaboy_ref.tex
%
%